\documentclass[11pt]{article} 
 
\begin{document} 
 
\title{Oil filaments produced by an impeller in a water stirred thank  \\ Fluid Dynamics Videos} 
 
\author{Rene Sanjuan-Galindo, Enrique Soto, Gabriel Ascanio and Roberto Zenit\\ 
\\\vspace{6pt} Universidad Nacional Autonoma de Mexico, Mexico, Distrito Federal, Mexico} 
 
\maketitle 
 
 
\begin{abstract}
 In this video (\href{http://ecommons.library.cornell.edu/bitstream/1813/8237/2/LIFTED_H2_EMS
T_FUEL.mpg}{Video 
1} and 
\href{http://ecommons.library.cornell.edu/bitstream/1813/8237/4/LIFTED_H2_IEM
_FUEL.mpg}{Video 
2}), the mechanism followed to disperse an oil phase in water using a Scaba impeller in a cylindrical tank is presented. Castor oil (viscosity = 500 mPas) is used and the Reynolds number was fixed to 24,000. The process was recorded with a high-speed camera. Initially, the oil is at the air water interface. At the beginning of the stirring, the oil is dragged into the liquid bulk and rotates around the impeller shaft, then is pushed radially into the flow ejected by the impeller. 
In this region, the flow is turbulent and exhibits velocity gradients that contribute to elongate the oil phase. Viscous thin filaments are generated and expelled from the impeller. Thereafter, the filaments are elongated and break to form drops. This process is repeated in all the oil phase and drops are incorporated into the dispersion. Two main zones can be identified in the tank: the impeller discharge characterized by high turbulence and the rest of the flow where low velocity gradients appear. In this region surface forces dominate the inertial ones, and drops became spheroidal.
\end{abstract}
 
 
%
\end{document}